\documentclass[twocolumn,showpacs,showkeys,preprintnumbers,amsmath,amssymb]{revtex4}
\usepackage{graphicx}
\usepackage{dcolumn}
\usepackage{bm}
\usepackage{subfig}

\usepackage{color}

\begin{document}

\preprint{APS/123-QED}

\title{Effect of gravity on clustering patterns and inertial particle attractors in Kinematic Simulation}
\author{M. Farhan}
\author{F. C. G. A. Nicolleau}
\email{corresponding author: F.Nicolleau@Sheffield.ac.uk}
\author{A. F. Nowakowski}
\affiliation{Sheffield Fluid Mechanics Group, Department of Mechanical Engineering, The University of Sheffield, Sheffield, United Kingdom}

\date{\today}


\begin{abstract}
In this paper, we study the clustering of inertial particles using a periodic kinematic simulation. The systematic Lagrangian tracking of particles makes it possible to identify the particles' clustering patterns for different values of particle inertia and drift velocity. The different cases are characterised by different pairs of Stokes number $St$ and Froude number $Fr$. For the present study $0\leq St \leq 1$ and {$0.4 \leq Fr \leq 1.4$}. The main focus is to identify and then quantify the clustering attractor - when it exists -
that is the set of points in the physical space where the particles settle when time goes to infinity.
Depending on gravity effect and inertia values, the Lagrangian attractor can have different dimensions varying from the initial three-dimensional space to two-dimensional layers and one-dimensional attractors that can be shifted from a horizontal to a vertical position.
\end{abstract}


\maketitle

\section{Introduction:}

Clustering could be defined as the propensity of an initially uniformly distributed cloud of particles to accumulate in some regions of physical space. This is an important phenomenon to understand in order to explore, identify and possibly monitor some natural or hand-made mixing processes such as those causing rain formation \cite{Falkovich-al-2002},
sediment transportation \cite{Pan-al-2011}, fuel mixing and combustion.

There are different ways to analyse particle clustering in turbulent flow and Direct Numerical Simulation (DNS) is the most widely used method (e.g. \cite{Cencini-al-2006,Saw-al-2012,Falkovich-Pumir-2004}). Particle clustering depends on both the flow conditions and the particle characteristics. Different flow conditions can lead to different clusters. The clustering mechanism would be different in the inertial or dissipation range of  turbulent flow \cite{Bec-al-2007}. In our paper we only study the effect of the scales in the inertial range and this is possible by using a synthetic model where forcing and dissipation are not needed to develop the inertial range. While considering particle characteristics, most of the studies on particle clustering have been conducted in the absence of external forces on particles but the effect of gravity(external force) was discussed in relation to cloud physics and rain formation in \cite{Falkovich-Pumir-2004,Woittiez-et-al-2008}.

 More recently, the effect of gravity on clustering mechanism has been further emphasized in \cite{Park-Lee-2014,Gustavsson-al-2014,Bec-et-al-2014}. In the present study, to observe the clustering pattern in the presence of gravity, the particles are initially uniformly distributed in the Kinematic Simulation (KS) flow. Though there is no particular difficulty in considering particles with different inertia in Kinematic Simulation, this study is limited to mono-dispersed seeding i.e. particles having the same inertia. Furthermore, the particles are considered small enough so that they neither affect the flow  nor interact with each other (one-way coupling). The positions of particles are monitored as a function of time and a Lagrangian attractor is observed for some cases. That is, the initially distributed cloud of particles will end in a set of loci that does not evolve any further. The particles move within that set of loci which defines the structure of the Lagrangian attractor and its dependence on $St$ and $Fr$ numbers is studied here.

We only consider attractors with integer dimensions (one-dimensional and two-dimensional structures) which are easy to identify. Different types of methods can be found in the literature to identify and then quantify particle clustering patterns, e.g.: correlation dimension \cite{Bec-al-2007}, radial distribution function RDF \cite{Saw-al-2012}, average-distance-to-nearest-neighbour method \cite{Park-Lee-2014}. The selection of a method is mainly based on the objective of the study. For example,
the RDF has the advantage of being directly related to the droplet collision rate. For the present work, the Box counting method BCM and the nearest-neighbour distance analysis are implemented to identify the integer dimensions of Lagrangian attractor in the presence of gravity.


The paper is organised as follows: in \S~\ref{method} we introduce the KS model,\ its notations and its parameters.
The different kinds of Lagrangian attractor are discussed and introduced in \S~\ref{resultsdisc}.
A quantitative analysis is conducted in \S~\ref{secquant}.
Section~\ref{seconcl} summarises our main conclusions.

\section{Kinematic simulation technique:}
\label{method}

Kinematic Simulation (KS) is a particular case of synthetic turbulence where the focus is on particle's trajectory at the expense of solving the Navier-Stokes equation. An analytical formula `synthetic flow' is used for the Eulerian flow field. The simplicity of the KS model excludes some features of real turbulent flow but capture the part of the physics which is required to perform Lagrangian particle tracking. Such is the idea with synthetic turbulence which retains less information than the whole flow, but tries to keep what is paramount for the Lagrangian story.

KS modelling has been successfully employed and validated \cite{Fung-al-1992,Elliott-Majda1996,Malik-Vassilicos1999}.
This kind of simulation is much less computing-time consuming than DNS which is important for the present study where we need to run many cases (about 400 cases for up to 1200 turnover time). Each case corresponds to a given $St$, $Fr$ and time and involves 15625 particles.

With synthetic simulations, one can develop models where turbulence ingredients and complexity can be added step by step helping to understand their respective importance. These synthetic models can be a useful complement to Direct Numerical Simulation. In particular, KS was instrumental in discriminating between the role of Lagrangian and Eulerian correlations for vertical diffusion in stratified and rotating flows \cite{Nicolleau-Yu-2007}. With KS it is also possible to play with the spectral law \cite{Nicolleau-Nowakowski-2011} and its consequences in terms of particle's dispersion. We also refer to the work of \cite{Malik-2014a,Malik-2014b} for a discussion on
how the work on KS can help to understand the sweeping effect on two-particle dispersion.

KS was first introduced as a way to understand particle dispersion rather than particles clustering but we propose here a work getting back to the main strength
of KS. That is to provide a coherent Lagrangian framework where some parameters (e.g. spectra \cite{Nicolleau-Nowakowski-2011}, waves \cite{Nicolleau-Yu-2007,Nicolleau-al-2012-ftac}, ...) can be studied in details posing the basis for a comparison with experiments.
Previous work \cite{Izermans-et-al-2010,Meneguz-Reeks-2011} particularly supports the use of KS for studying the evolution of the particle cloud in the absence of gravity effect which made the study more about segregation than clustering.
\\[2ex]
As we are not interested in two-particle dispersion,
we limit our study to small Reynolds numbers, more precisely to scale ratio ${k_i}_{max}/{k_i}_{min}=15$\footnote{$i=1$, 2 or 3}.

\subsection{Periodic KS method for isotropic turbulence}

In Kinematic Simulation the underlying Eulerian velocity field is generated as a sum of random incompressible Fourier modes with a prescribed energy spectrum $E(k)$.
Here we limit the study to a Kolmogorov type spectral law $E(k) \sim k^{-5/3}$.
Using KS, the computational task reduces to calculate the trajectory of each particle placed in the turbulent field initially at $\mathbf{X}_0$.
Each trajectory is, for a given initial condition, solution of the differential equation set:
\begin{eqnarray}
{d\mathbf{X} \over dt} &=& \mathbf{V}(t)
\\
{d\mathbf{V} \over dt} &=& \mathfrak{F}(\mathbf{u}_{E}({\mathbf{X},t),\mathbf{V},t})
\label{a1}
\end{eqnarray}
where $\mathbf{X}(t)$ is the particle's position, $\mathbf{V}(t)$ its Lagrangian velocity and
${\bf u}_{E}$ the analytical Eulerian velocity used in KS. $\mathfrak{F}$ is a function relating the Lagrangian acceleration to the Eulerian and Lagrangian velocities.

In KS ${\bf u}_{E}$ takes the form of a truncated Fourier series, sum of $N_k=N^3$ Fourier modes:
\begin{equation}
\mathbf{u}(\mathbf{x})=\sum_{i=1}^{N}\sum_{j=1}^{N}\sum_{K=1}^{N}\mathbf{a_{ijl}}\rm{cos}(\mathbf{k_{ijl}}.\mathbf{x})+\mathbf{b_{ijl}}\rm{sin}(\mathbf{k_{ijl}}.\mathbf{x})
\label{EqKSfield}
\end{equation}
where
$\mathbf{a_{ijl}}$ and $\mathbf{b_{ijl}}$ are the decomposition coefficients corresponding to the wavevector $\mathbf{k_{ijl}}$.
In its general form the KS field can also be a function of time but we limit the study to a steady KS. The effect of introducing a time-dependence in the Fourier modes will be the objective of future study.

\begin{figure*}[!ht]
\includegraphics [width=1.\columnwidth]{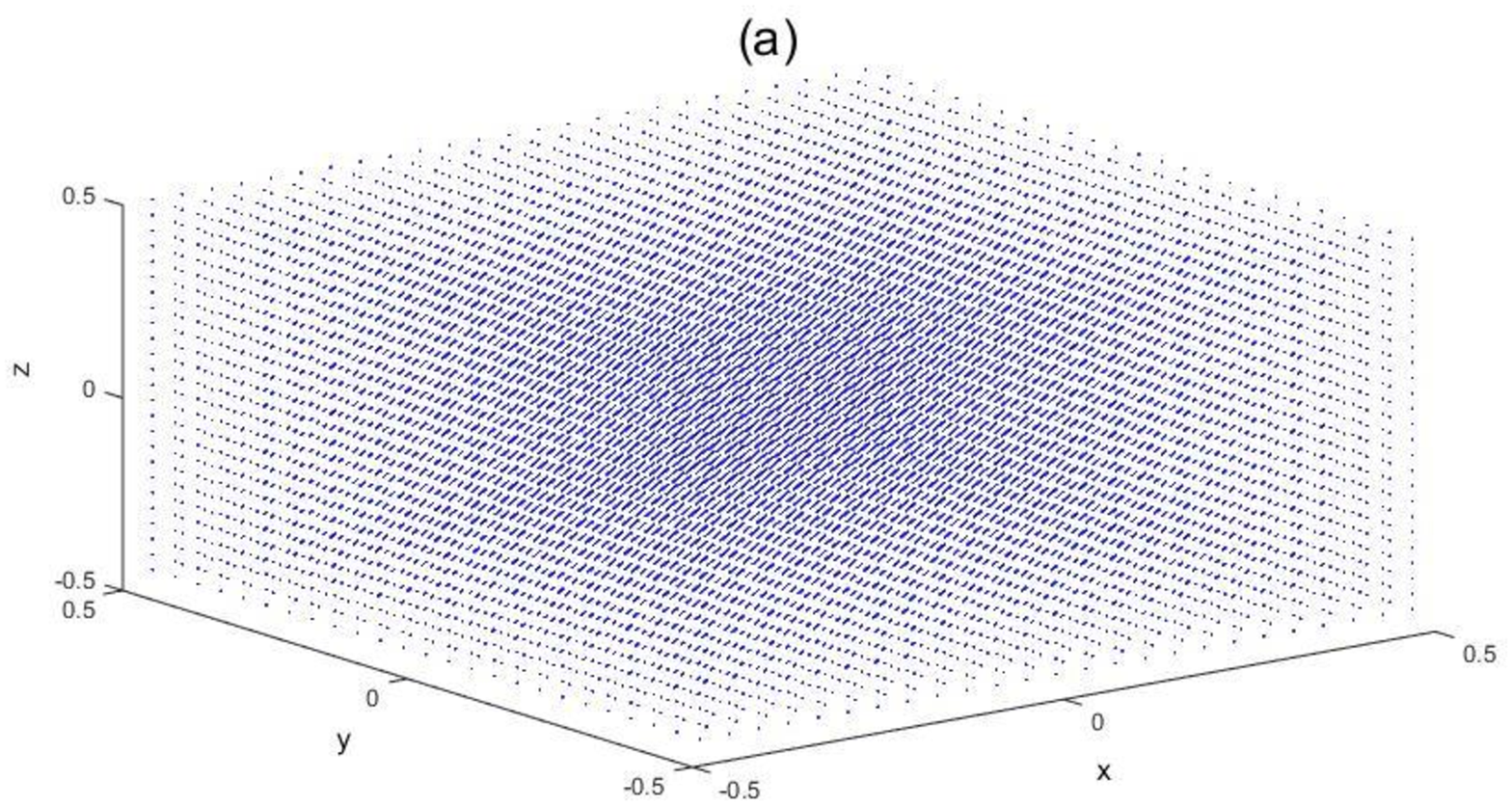}
\includegraphics [scale=0.4]{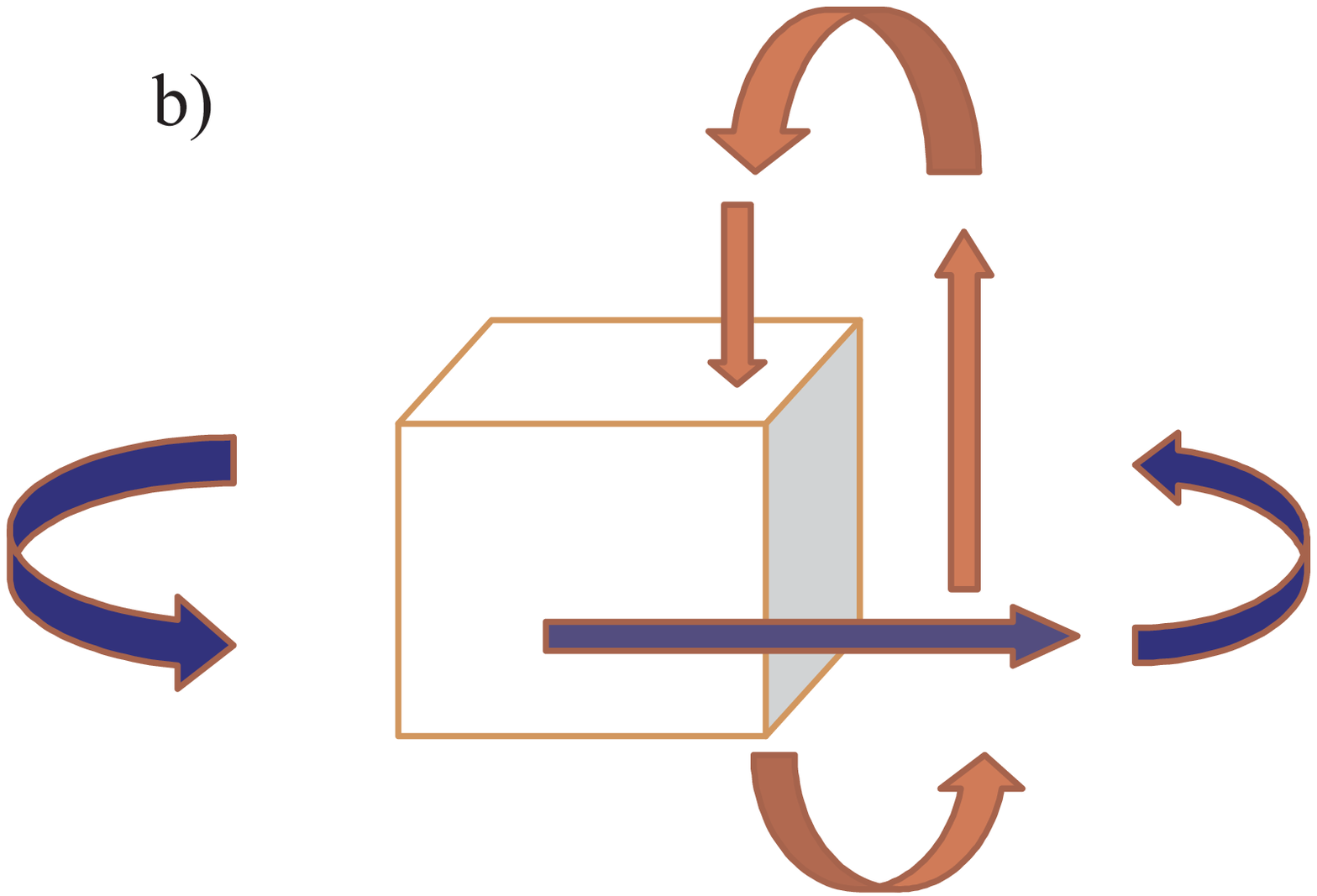}
\caption{\label{fig-01-partpos} Particles' initial distribution (a) and re-injection (b)}
\end{figure*}

Unlike the classic KS decomposition \citep{Fung-Vassilicos-1998,Nicolleau-ElMaihy-2006}, here the wavevectors $\mathbf{k_{ijl}}=(k_{i},k_j,k_l)$
are implemented arithmetically to enforce a periodic condition for the flow field:
\begin{eqnarray}
k_{i} &=& \frac{2 \pi}{L_{x}} (n_{i}-1)
\label{eqn: geometrical decimation1}
\\
k_{j} &=& \frac{2 \pi}{L_{y}} (n_{j}-1)
\label{eqn: geometrical decimation2}
\\
k_{l} &=& \frac{2 \pi}{L_{z}} (n_{l}-1)
\label{eqn: geometrical decimation3}
\end{eqnarray}
where $(n_{i},n_{j},n_{k})$ are integers satisfying $1 \le n_{i} \le N$.
In practice, we choose $(L_{x}=L_{y}=L_{z})$ for creating isotropic turbulence and to ensure the flow incompressibility the Fourier coefficient vectors $\mathbf{a_{ijl}}$ and $\mathbf{b_{ijl}}$ are set orthogonal to the wavevector:
\begin{equation}
\mathbf{a_{ijl} \cdot k_{ijl}}=\mathbf{b_{ijl} \cdot k_{ijl}}=0
\end{equation}
Their magnitude is fixed by the energy spectrum, $E(k)$
\begin{equation}
\left|\mathbf{a_{ijk}}\right|^{2}= \left|\mathbf{b_{ijk}}\right|^{2} = 2E(k)\Delta k_{ijl} / m_k
\end{equation}
where $m_k$ is the number of wavevectors of wavenumber $k=\|\mathbf{k_{ijl}}\|$.
The spectrum follows the Kolmogorov form in the inertial range,
\begin{equation}
E(k)= A \, k^{-5/3} \mbox{\, for \,} k_{min} \le k \le k_{max}
\label{eqn:inertial range spectrum}
\end{equation}
where $A$ is a constant. From the spectral law, the rms velocity and the integral length scale can be defined as follows:
\begin{equation}
u_{rms} = \sqrt{ {2 \over 3} \int_{k_{min}}^{k_{max}} E(k) dk}
\label{a8}
\end{equation}
\begin{equation}
{\mathcal{L} = { 3 \pi \over 4} { \int_{k_{min}}^{k_{max}} k^{-1} E(k) dk \over \int_{k_{min}}^{k_{max}} E(k) dk }}
\label{a9}
\end{equation}
The Kolmogorov length scale is defined as $ \eta = 2\pi/k_{max}$,
whereas the largest physical scale is $L= 2\pi / k_{min}$ which determines the inertial range
$[\eta,L]$ over which (\ref{eqn:inertial range spectrum}) is observed.
It is worth noting that $\mathcal{L} \simeq L$ for sufficiently large inertial ranges.
However, here in contrast to other KS studies the inertial range is small and  $L \simeq 5 \mathcal{L}$. In this paper, non-dimensional numbers ($St$ and $Fr$) are
based on the integral length scale $\mathcal{L}$ and for the sake of future comparisons both are reported in Table~\ref{tabKS}. The ratio between the largest length scale and the Kolmogorov length scale is $k_{max} / k_{min}$
and the associated Reynolds number is: $Re_{{L}} = (k_{max}/k_{min})^{4/3}$.
This is the standard way to define a Reynolds number in KS and a DNS or an experiment
yielding the same ratio $k_{max}/k_{min}$ would have a much larger Reynolds number. Finally, a characteristic time for normalisation can be $t_d = {L}/u_{rms}$ or $\mathcal{T}= \mathcal{L}/u_{rms}$. All the periodic KS parameters are gathered in Table~\ref{tabKS}.
\begin{table}[h]
\caption{\label{tabKS} Periodic KS parameters}
\newcolumntype{d}[1]{D{.}{\cdot}{#1}}
\begin{tabular}{ld{-1}}
\hline
\hline
$L_x=L_y=L_z$ & 1
\\
$N$ & 10
\\
$N_p$ & 15625
\\
$u_{rms}$ & 0.8703
\\
$\mathcal{L}$ & 0.2106
\\
$L$ & 1
\\
$\eta$ & 0.0642
\\
$\mathcal{T}$ & 0.2420
\\
$t_d$ & 1.1491
\\
${k_i}_/{k_i}_{min}$ & 15
\\
$k_{max}/k_{min}$ & 15.5885
\\
$Re_L$ & 38.94
\\
\hline
\hline
\end{tabular}
\end{table}
\\
The particles are initially homogeneously distributed as shown in Fig.~\ref{fig-01-partpos}a and whenever a particle leaves the turbulence box domain
(e.g. $\textbf{X}_i >L_x$), then it is re-injected from the opposite side as shown in Fig.~\ref{fig-01-partpos}b to keep the periodic condition.


\subsection{Equation of motion}

Following \cite{El-Azm-Nicolleau-2008} the equation of motion for the inertial particle is derived from \cite{Gatignol-1983,Maxey-Riley-1983} and consists of a drag force and drift acceleration (weight):
\begin{equation}
{d{\bf V} \over dt} = \frac{1}{\tau_{a}} \left ( {\bf u}({\bf x}_{p}(t),t)-{\bf V}(t)+{\bf V}_{d} \right )
\label{E2-29}
\end{equation}

\noindent where $\tau_{a}$ is the particle's aerodynamic response time and $V_{d} = \tau_{a} \; {\bf g}$ the particle's terminal fall velocity or drift velocity.

\subsection{Non-dimensional parameters}

Three non-dimensional parameters are introduced to make qualitative and quantitative analyses of the particle clustering.

\begin{itemize}
\item
 The Stokes number expresses the ratio between the particle's response time (inertia effect) and the turbulence characteristic time
    \begin{equation}
        St= \tau_{a}/\mathcal{T} = \tau_{a} u_{rms} / \mathcal{L}
        \label{E2-30}
    \end{equation}
 It measures the relative importance of the particle inertia. In the limiting case $St = 0$; the heavy particles recover the motion of the fluid tracers, whereas for $St \rightarrow \infty $ the heavy particles become less and less influenced by the surrounding velocity field.
%
\item
The Froude number is the ratio between inertial forces and gravitational forces.
    \begin{equation}
        Fr = u_{rms}/ \sqrt{g\mathcal{L}}
        \label{E2-31b}
    \end{equation}
In our study the rms velocity $u_{rms}$ and inertial length scale are constant and $g$ is varied.
\item
The Drift parameter is the ratio between the particle's drift velocity and the turbulence rms velocity:
    \begin{equation}
        \gamma = V_{d} / u_{rms}
        \label{E2-31}
    \end{equation}
The Drift parameter can still be defined without gravity. Then
$\gamma$ can be considered as measuring the
effect of a mean velocity $V_d$.

If $V_d$ is caused by gravity:
    \begin{equation}
        \gamma = \tau_{a}g / u_{rms}
        \label{E2-31b}
    \end{equation}
in this case the drift parameter is affected by both the gravity and the particle's inertia.
\item $\gamma$ can be expressed as a function of Stokes and Froude numbers so for a given turbulence the case corresponding to a constant gravity, that is varying $\tau_a$ only, is given by
\begin{eqnarray}
Fr & = & \mbox{constant}
\\
\gamma &\sim & St
\label{E2-300}
\end{eqnarray}
\end{itemize}

\section{Results and Discussion}
\label{resultsdisc}
\subsection{Clustering Pattern variations in relation to time of evolution}

\begin{figure*}[!ht]
\includegraphics [scale=0.3] {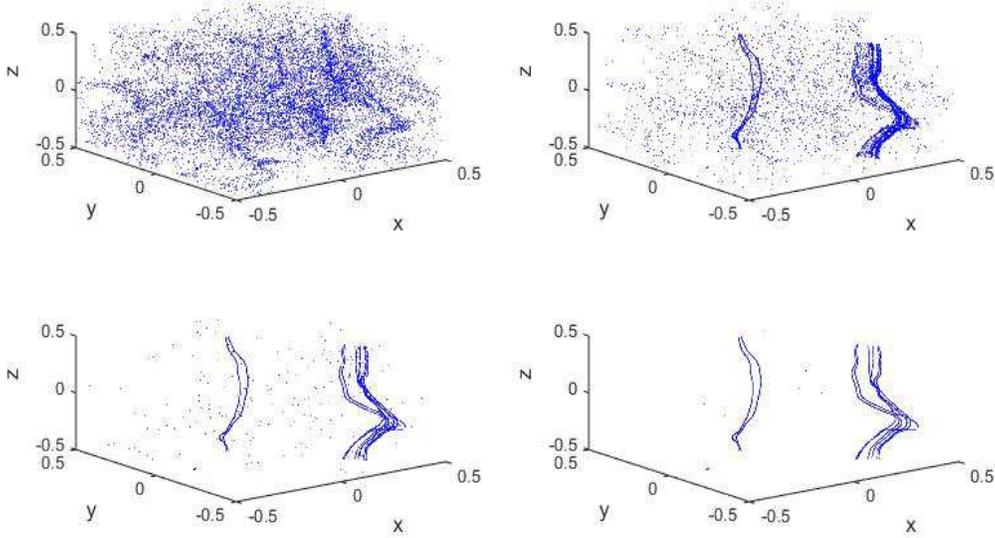}
\caption{\label{Fig2} Time evolution of inertial particles for $St=0.207$ and $Fr=0.55$ with top-left t=1, top-right t=10, bottom-left t=600 and bottom-right t=1200}
\end{figure*}

\begin{figure*}[!ht]
\includegraphics [scale=0.3] {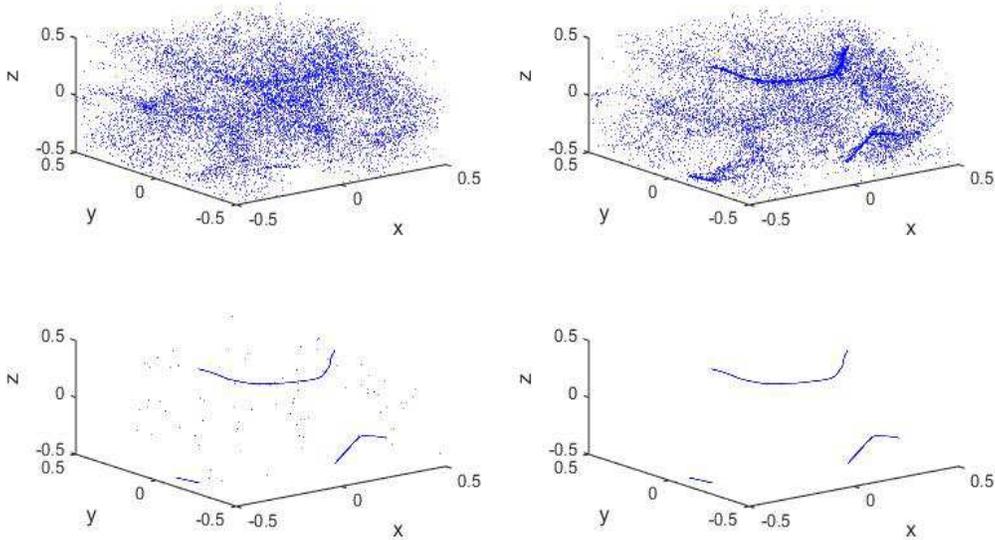}
\caption{\label{Fig3} Time evolution of inertial particles for $St=0.413$ and $Fr=0.85$ with top-left t=1, top-right t=5, bottom-left t=100 and bottom-right t=300 }
\end{figure*}

The particles initially uniformly distributed in the flow field are allowed to evolve until a Lagrangian attractor is achieved. The shape of the attractor varies from clear one-dimensional structures (Figs~\ref{Fig2} and \ref{Fig3})
to three-dimension distributed structures (Fig.~\ref{Fig4}de) or two-dimensional curtain-like structures (Fig.~\ref{Fig4}dh). For a short time the attractor's shape is time-dependent as shown in Figs~\ref{Fig2} and \ref{Fig3}. The time evolution of the cluster depends on non-dimensional parameters $St$ and $Fr$ as illustrated in Fig.~\ref{Fig2} where it takes 4 times longer to reach the one-dimensional Lagrangian attractor than in the case of Fig.~\ref{Fig3}. In this paper, we do not intend to study the temporal evolution of cluster attractors and  are only interested in attractor's asymptotic form (i.e. for $ t \to \infty$).

In this section, we focus on the qualitative measure of attractors and only a few cases are presented in Fig.\ref{Fig4} and Fig\ref{Fig11}. A systematic quantification will be proposed in \S~\ref{secquant} which consists of a comprehensive set of data generated with small increments in $Fr$ and $St$ numbers. As the one-dimensional Lagrangian attractor is observed for various pairs of $St$ and $Fr$, it was difficult to suggest any definite relationship between one-dimensional clustering and $St$ and $Fr$ . Hence, a scheme of further classification is adopted to establish such a relationship.
We use the following nomenclature:
\begin{itemize}
\item[i)] 1D-H : horizontal one-dimensional Lagrangian attractor as in Fig.~\ref{Fig3}bf,
\item[ii)] 1D-V : vertical one-dimensional Lagrangian attractor as in Fig.~\ref{Fig4}bg,
\item[iii)] 1D-HV : Intermediate one-dimensional Lagrangian attractor as in Fig.~\ref{Fig4}ag,cg,
\item[iv)] 2D-L : two-dimensional vertical curtain-like layer as in Fig.~\ref{Fig4}dh (see also \cite{Woittiez-et-al-2008}),
\item[v)] 3D : any three-dimensional structure without any particular structure in the cloud as in Fig.~\ref{Fig4}de.
\end{itemize}

The qualitative results are split into three different categories which can take into account the effect of gravity and/or inertia:
\begin{itemize}
\item[i)] keeping $St$ constant \S\ref{secSt},
\item[ii)] keeping $Fr$ constant \S\ref{secFr},
\item[iii)] keeping $\gamma$ constant \S\ref{secgamma}.
\end{itemize}


\subsection{\label{secSt}Clustering Pattern variations in relation to Constant Stokes Number $St$}

\begin{figure*}[!ht]
\includegraphics [scale=0.25]{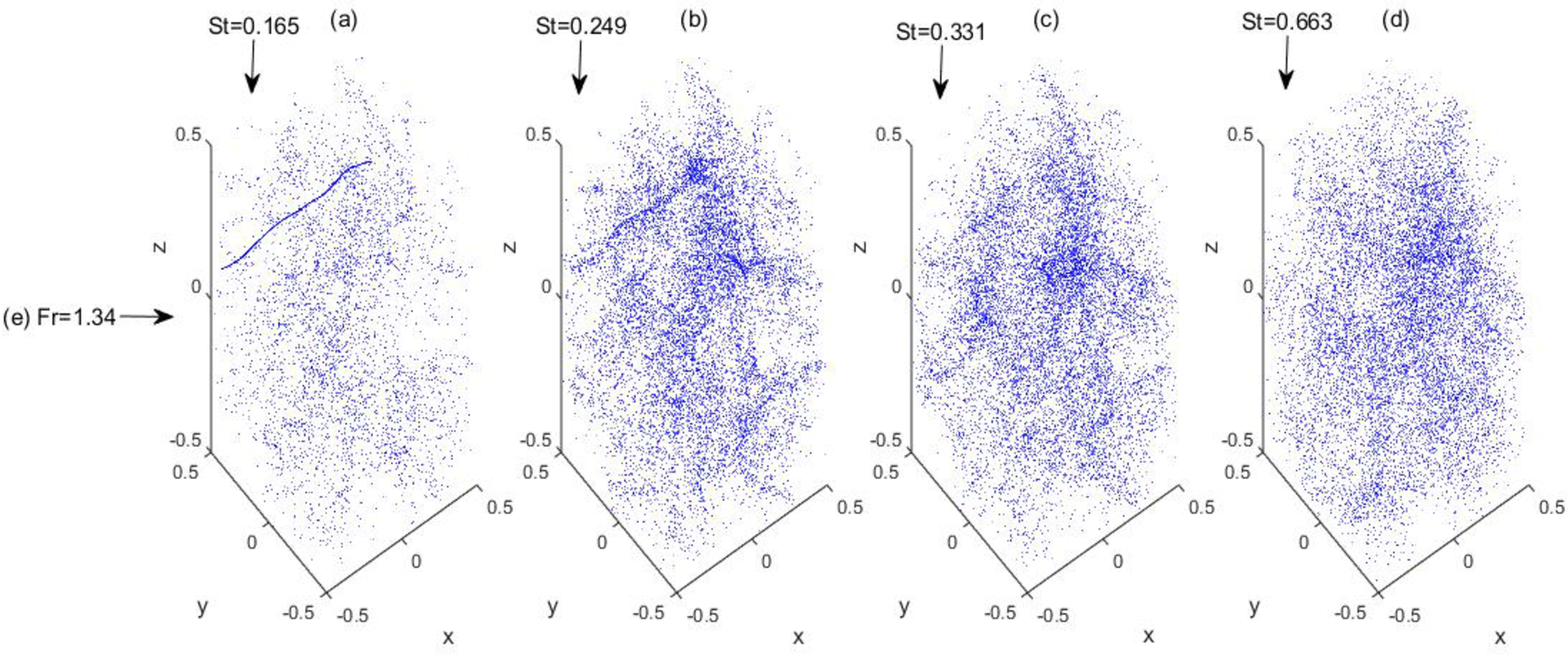}
\includegraphics [scale=0.25]{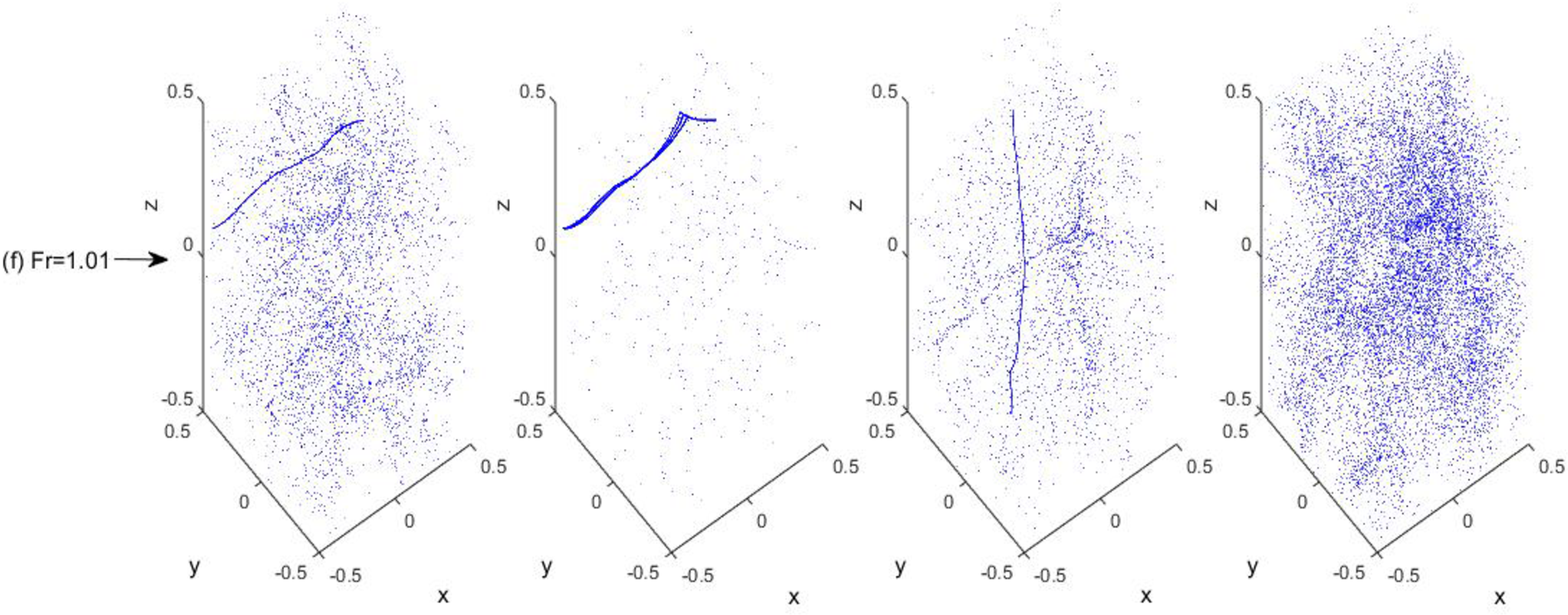}
\includegraphics [scale=0.25]{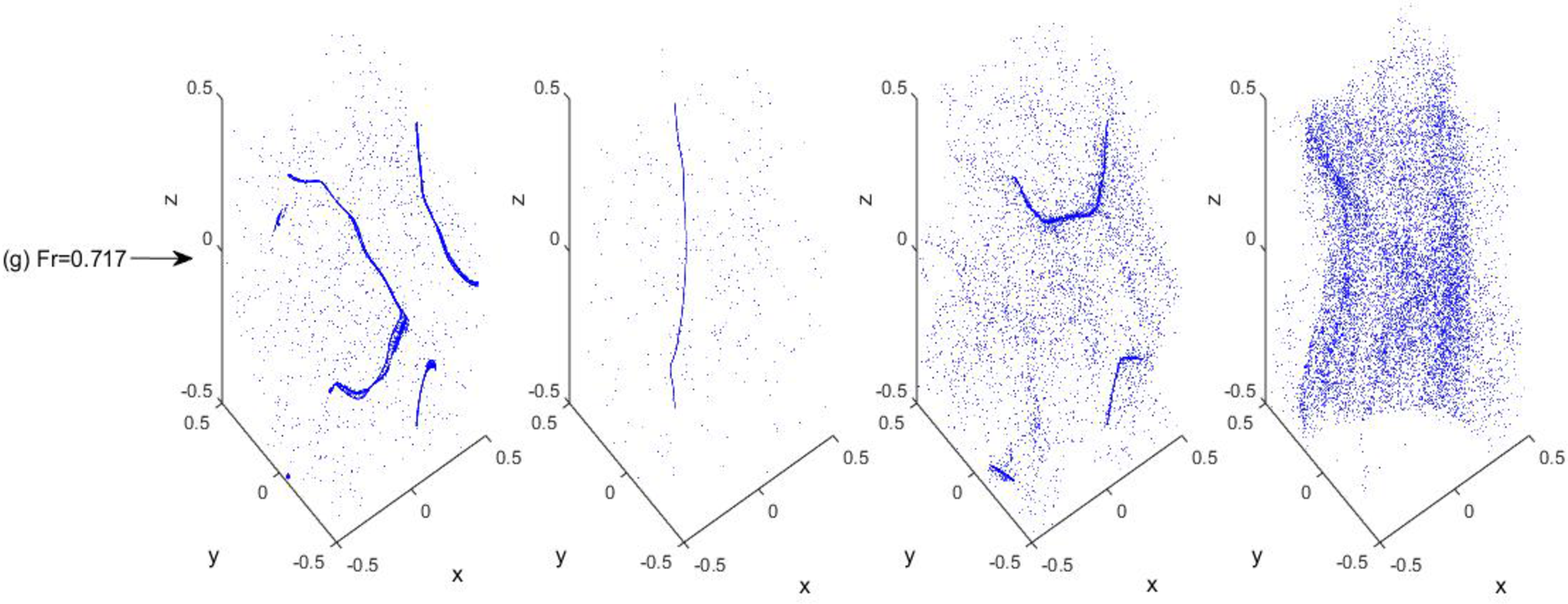}
\includegraphics [scale=0.25]{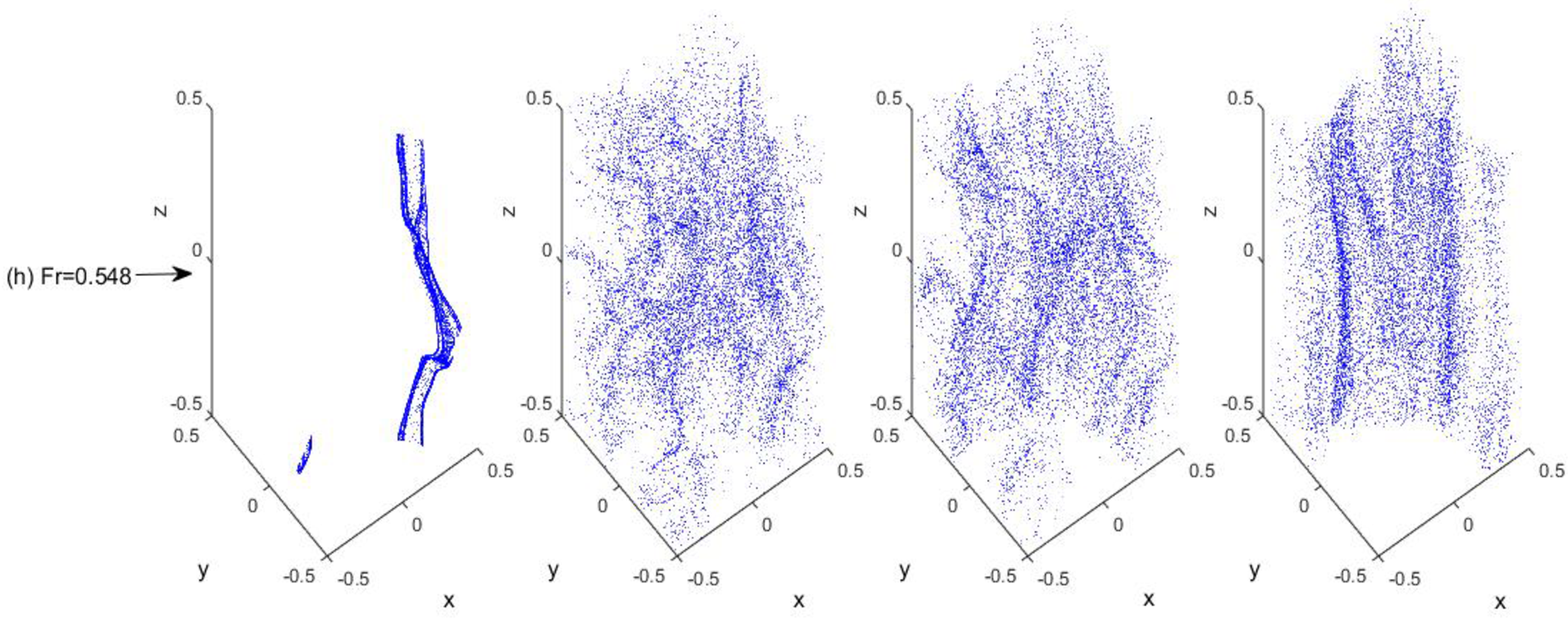}
\caption{\label{Fig4}Evolution of the particles cloud for $0.548\le Fr \le 1.34$ and $0.663\le St \le 1.165$, at $t=300$}
\end{figure*}

We can analyse the results by fixing the Stokes number $St$ and varying the Froude number $Fr$.
Four representative cases with varying values of $Fr$ are shown in Fig.~\ref{Fig4}, in columns a, b, c and d, namely $St=0.165$, 0.249, 0.331 and 0.663. For each of the Stokes numbers, we explored within the range $0.548\leq Fr \leq 1.34$.

 As the Froude number decreases, the particles first cluster on a one-dimensional Lagrangian attractor then they may rearrange on another 1D or 2D Lagrangian attractor (Fig.~\ref{Fig4} bh, ch, dg and dh). The Lagrangian attractor also moves from a predominantly horizontal (Fig.~\ref{Fig4} ae, af and bf) to a vertical direction (Fig.~\ref{Fig4} ag, bg and cf) with decreasing value of $Fr$.

The qualitative shapes with varying values of $Fr$ are listed in Table~\ref{Table-1}. When one-dimensional structures are in-between vertical and horizontal (1D-HV) directions as in Fig.~\ref{Fig4}bf we ticked both 1D-H and 1D-V in the table.
\begin{table}[htb]
\caption{\label{Table-1}Different cases for studying the attractor topology for different ranges of $St$.}
\begin{center}
\begin{tabular}{lllcccc}
\hline
\hline
Case & $St$ range & $Fr$ range & \multicolumn{4}{c}{observed patterns}
\\
 & & & 1D-H & 1D-V & 2D-L & 3D
\\
\hline \hline
\\
A& 0.041-0.124 & 0.42-1.34 & \checkmark & & & \checkmark
\\
B& 0.165-0.300 & 0.42-1.34 & \checkmark & \checkmark & &\checkmark
\\
C& 0.331-0.413 & 0.42-1.34& & \checkmark &\checkmark & \checkmark
\\
D& 0.496-1.000 & 0.42-1.34 & & & \checkmark & \checkmark
\\
\hline
\hline
\end{tabular}
\end{center}
\end{table}

Now we describe each case one by one according to observed patterns. For Case A, corresponding to low values of $St$, the decrease in $Fr$ (increase in gravity) forces the particles to cluster in a horizontal direction whereas the particles will disperse evenly as $Fr$ is further decreased. The increases in gravity destroy the particles' clustering for low range of $St$. Similarly, for case B, initially distributed inertial particles cluster into a horizontal attractor (1D-H) Fig.~\ref{Fig4}bf and then an increase in the gravity effect (lower $Fr$) causes the particles to accumulate in the vertical direction (1D-V) as in Fig.~\ref{Fig4}bg. With further increases in the gravity they recover a 3D distribution as shown in Fig.~\ref{Fig4}bh.The appearance of 1D-V in case B shows the greater effectiveness of gravity at a relatively high range of the $St$ for a given value of $Fr$.

For further increases in the Stokes number $St$, case C, the horizontal structure (1D-H) is not observed. Rather a vertical 1D pattern (1D-V) is seen as in Fig.~\ref{Fig4}cf  which can transform into a 1D-HV  attractor as in Fig.~\ref{Fig4}cg with decreasing values of $Fr$. This implies that the particle inertia starts dominating over the flow Eulerian structure and allows the gravity to play a more important role. At higher values of $St$ (case D), there is no more one-dimensional clustering but some clustering can still be observed in the form of two-dimensional vertical curtain-like structures as shown in Fig.~\ref{Fig4}dh at low values of $Fr$.

\subsection{\label{secFr}Clustering Pattern variations in relation to Constant Froude Number $Fr$}

A constant Froude number corresponds to the case of varying the particle's property ($\tau_a$) for a given environment (turbulence and gravity) which exists in most of the experimental situations. The variations in clustering patterns are identified by keeping $Fr$ constant while varying $St$. For the purpose of qualitative measures, three different cases as shown in Table~\ref{Table-2} are considered with small increments in the $St$ ranging [0-1].
\begin{table}[htb]
\caption{\label{Table-2}Different cases for studying the attractor topology for different ranges of $Fr$.}
\begin{center}
\begin{tabular}{llccccc}
\hline \hline
Case & $Fr$& $St$ range & \multicolumn{4}{c}{observed patterns}
\\
 & & & 1D-H & 1D-V & 2D-L & 3D
\\
\hline \hline
\\
E& 1.01 & 0-1 & \checkmark & \checkmark & & \checkmark
\\
F& 0.717 &0-1 & & \checkmark & \checkmark & \checkmark
\\
G& 0.548 &0-1 & & \checkmark & \checkmark &
\\
\hline \hline
\end{tabular}
\end{center}
\end{table}
\\
In Fig.~\ref{Fig4} cases of constant Fr correspond to the horizontal rows e, f, g, h. As $St$ increases, the particles' one-dimensional clustering is first enhanced and then destroyed to eventually reappear in the form of a two-dimensional layer (2D-L).

For high values of $Fr$ (low gravity), case E corresponding to Fig.~\ref{Fig4}f, particles settled on horizontal one-dimensional structures (1D-H) for low values of $St$. The increase in $St$ values resulted into vertical one-dimensional structures (1D-V). For the mid-range values of $Fr$, case F corresponding to Fig.~\ref{Fig4}g, the clear one-dimensional horizontal structure (1D-H) is no longer observed but instead some intermediate (1D-HV) one-dimensional structures can be seen for low $St$ values which converge into a layered curtain-like (2D-L) structure as $St$ is increased. Finally, low values of $Fr$ (case G, Fig.~\ref{Fig4}h) allow the particles to accumulate predominantly in the direction of gravity, so vertical patterns are identified such as 1D-V and 2D-L structures.

\subsection{\label{secgamma}Clustering Pattern variations in relation to constant drift parameter $\gamma$}

\begin{figure*}[!ht]
   \includegraphics [scale=0.23]{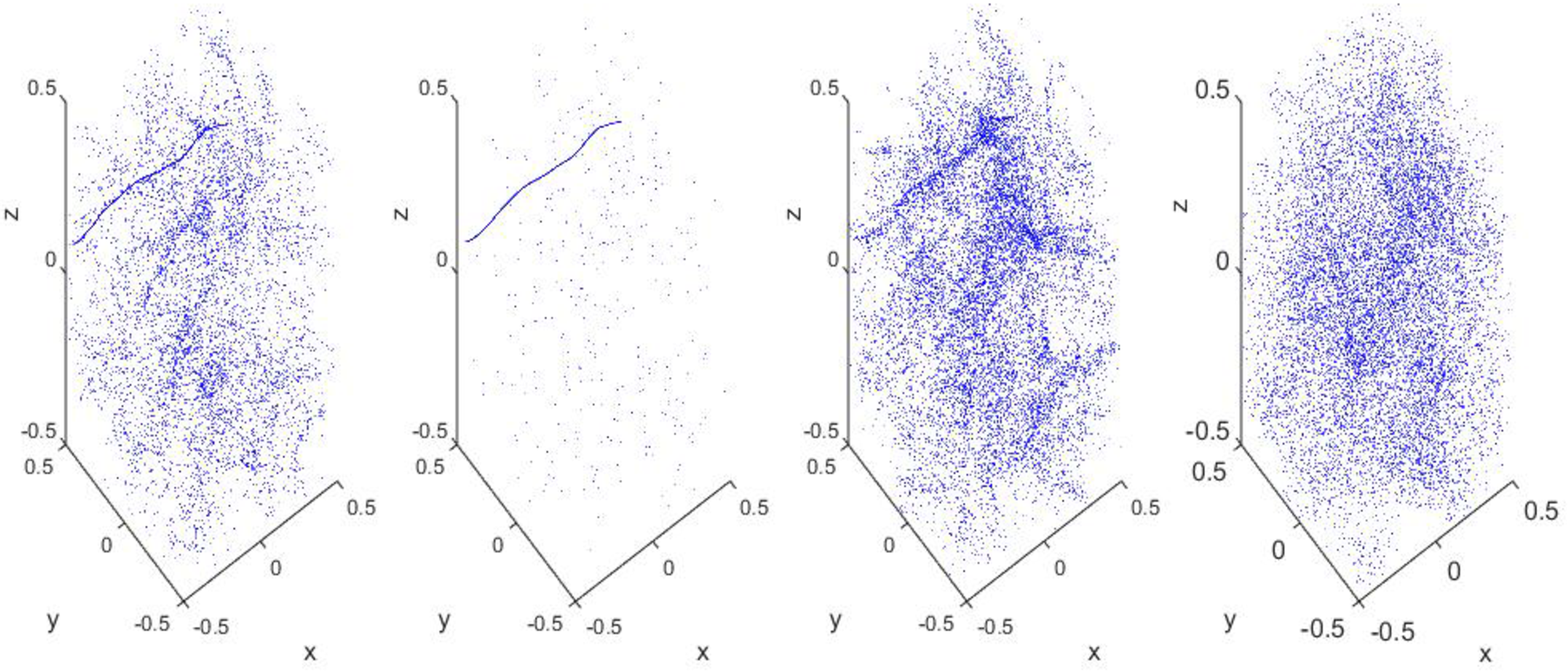}

   \includegraphics [scale=0.23]{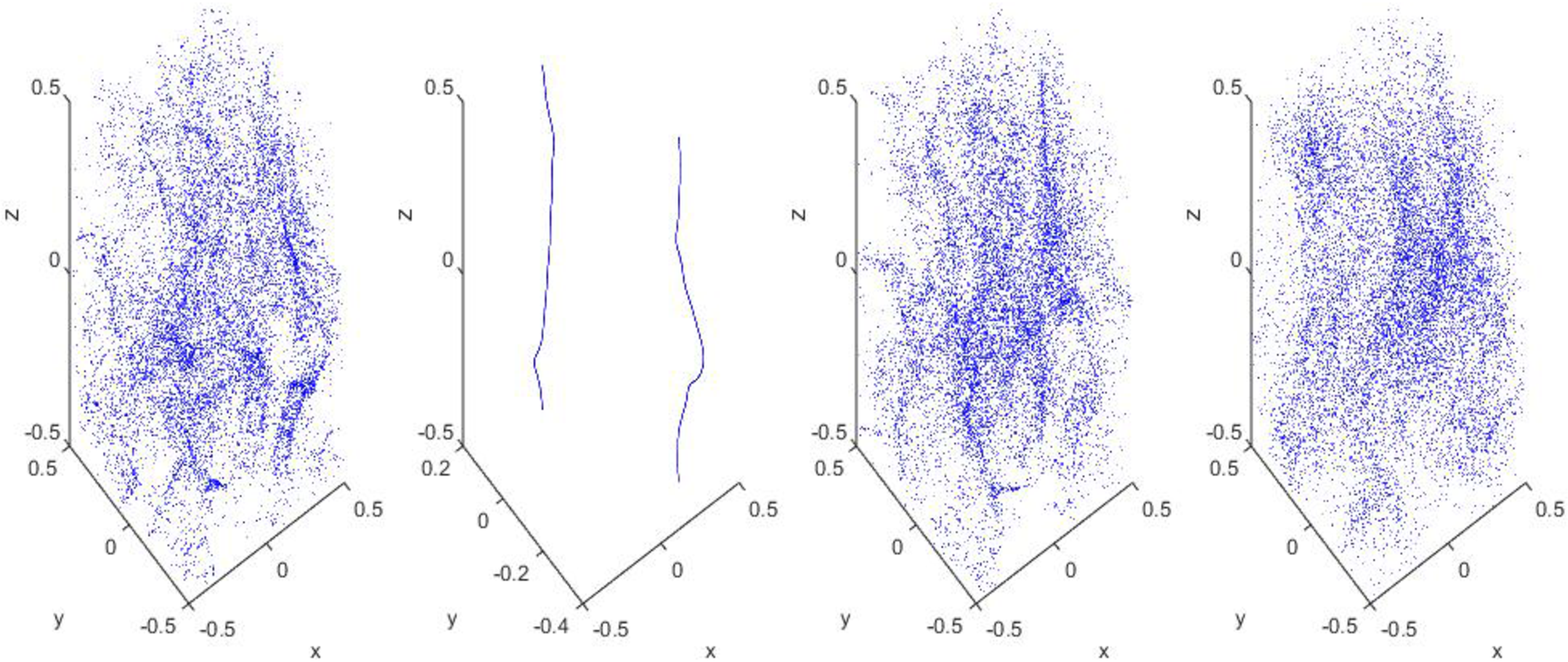}
   
   \includegraphics [scale=0.23]{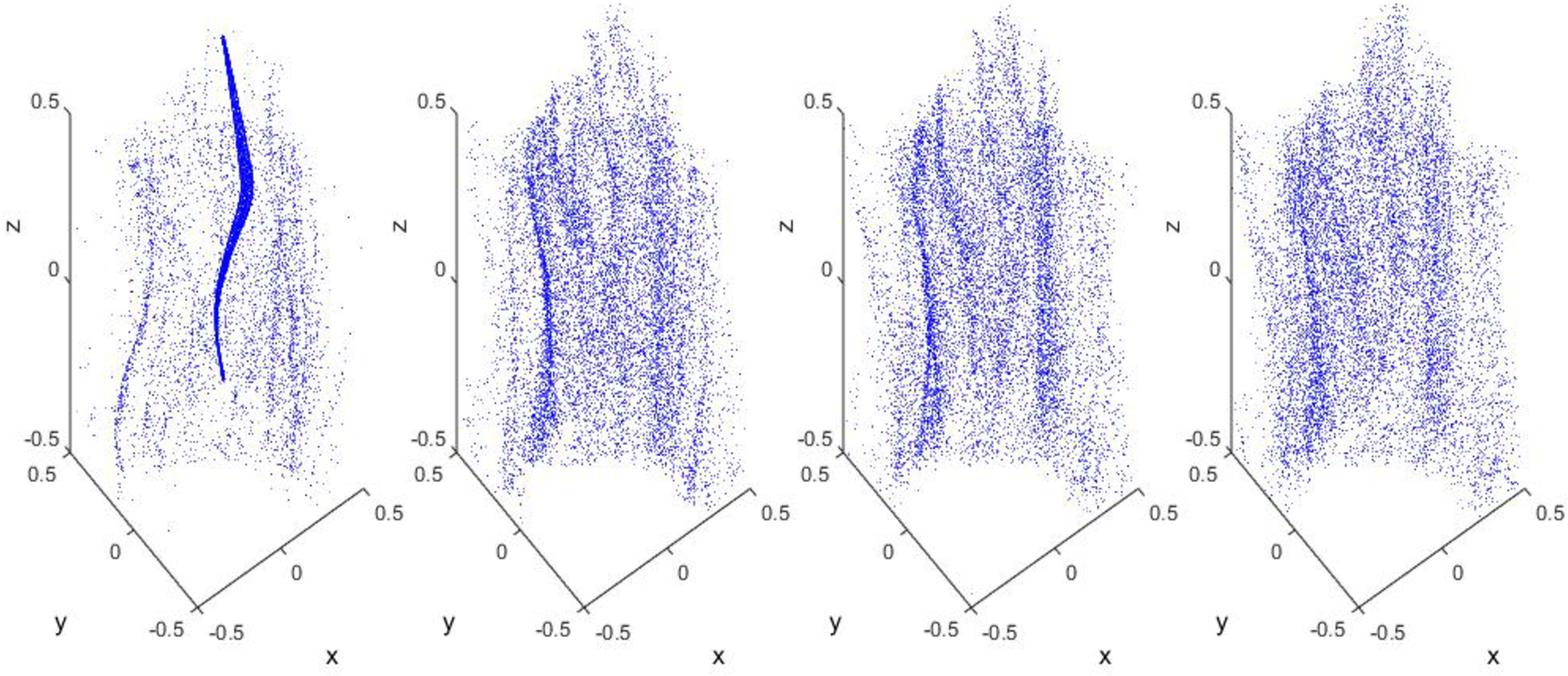}
\caption{\label{Fig11}Evolution of particles with increasing $St$ for a given $\gamma$ at $t=300$.
Top $\gamma=0.138$, left to right St= 0.100, 0.165, 0.249 and 1; middle $\gamma=0.689$, left to right St= 0.124, 0.165, 0.249 and 0.827; bottom $\gamma=2$, left to right St= 0.207, 0.600, 0.827 and 1.}
\end{figure*}

It results from the previous discussion that the variations in inertial and gravity effects do not have a monotonic effect on the particle clustering. Physically gravity and inertia are combined effects but one can consider a particle subjected to a drift velocity without referring explicitly to gravity. This effect of drift can be assessed by identifying the patterns with the drift parameter $\gamma$ instead of $Fr$. So here we want to observe the variation in the particle attractor by keeping the drift parameter $\gamma$ constant.
Fig.~\ref{Fig11} shows the three cases $\gamma=0.135$, 0.689 and 2, as $St$ increases for the arbitrary time $t=300$. It is not possible to keep the same range for $St$ for all cases because $St$ and $\gamma$ are linked.

For low values of $0 \leq \gamma \leq 0.2$ (Fig.~\ref{Fig11} top), particles tend to accumulate as a horizontal attractor for low values of $St$. As observed previously, further increases in $St$ lead to particles scattering. Eventually particles disperse evenly  in the flow for high values of $St$.

As $\gamma$ increases $0.2\leq \gamma\leq 0.8$ (Fig.~\ref{Fig11} middle), horizontal attractors are not observed any more for low values of $St$, instead some intermediate 1D-HV and vertical 1D-V attractors are observed with increasing values of $St$. Further increases in inertia disperse the particles evenly in the flow field.
Third case $0.8<\gamma<2$ (Fig.~\ref{Fig11} bottom) corresponds to relatively high values of $\gamma$. The particles are trapped in a 2D-L structure.
The increase in $St$ results into the particles dispersing more homogeneously on this 2D-L attractor.

To summarise: an increase in $\gamma$ can lead to a 1D-V or 2D-L rather than a 1D-H attractor and an increase in $St$ destroys the one dimensional attractor leading to the particles reorganising on a 2D-L.
\\
This is in agreement with \cite{Park-Lee-2014} who mention gravity-driven clustering of inertial particles in turbulence
and report a different kind of particle clustering caused purely by gravity, that is, clustering in a vertical stripe pattern formed when strong gravity acts on heavy particles.

\section{\label{secquant}Quantification of Clustering Patterns}

Visualizations of the particle cloud for small discrete increments of the non dimensional parameters $St$ and $Fr$ can be tedious. It means looking at about 400 cases in this study in a systematic order. Beyond the simple visualisation, it is important to quantify the Lagrangian attractors using an appropriate method for spatial clustering. Two different methods are considers here: the box-counting method (BCM) and the average-distance-to-nearest-neighbour $\Delta$. The average-distance-to-nearest-neighbour is eventually chosen for the final quantitative analysis.

\subsection{Box Counting Method}

\begin{figure}[!ht]
   \includegraphics [width=1.\columnwidth,height=0.75\columnwidth]{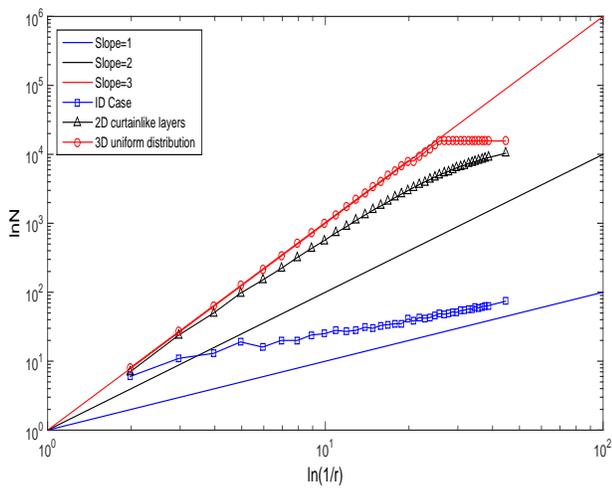}
\caption{\label{Fig14} (Color
  online) Bench mark for BCM }
\end{figure}

\begin{figure}[!ht]
\includegraphics [width=1.\columnwidth,height=0.75\columnwidth]{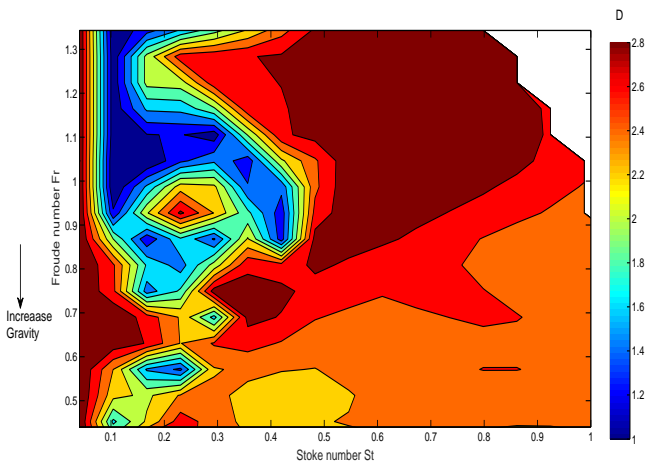}
\caption{\label{Fig15} (Color
  online) Contour plot of the attractor fractal dimension $D$ as a function of $(St, Fr)$.}
\end{figure}

The box counting method (BCM) is a commonly used method to determine the fractal dimension of an object. Though in our simulation the range of scales is too short to observe the fractal patterns described in \cite{Nicolleau-ElMaihy-2004}, BCM remains a useful tool to discriminate between
one-dimension, two-dimension and three-dimension clustering patterns.
The fractal dimension, $D$, represents the relation between the box size, $r$ and $N_r$ the number of boxes needed to cover the cloud of particles, that is:
\begin{equation}
N_r \sim r^{D}
\label{E4-1}
\end{equation}
It is straightforward to obtain the fractal dimension from a log-log plot:
\begin{equation}
D = \frac{\Delta\ln {N_r}}{\Delta \ln {r}}
\label{E4-2}
\end{equation}
A validation of the method is made on three clear identified shapes, namely the one-dimensional Lagrangian attractor, the two-dimensional curtain-like layered pattern and a three dimensional distribution. As shown in Fig.~\ref{Fig14}, the difference between these three patterns is clearly captured.
\\[2ex]
The BCM is sensitive to the achievement of the attractor, that is if few particles have not settled on the attractor they can alter the box counting results. So, with this method it is necessary to make sure that the cloud shape is the asymptotic final one which requires a very long time. As illustrated in Fig.~\ref{Fig2}: though at a very short time $t=10$ the position and shape of the 1D-V is obvious, it is necessary to wait up to $t=1200$ to get the final cluster position that will allow a correct measure for the BCM. All the cases are reported  in Fig.~\ref{Fig15} for clusters having reached their final shapes (attractors). Iso-contours of $D$ as a function of $(St, Fr)$ are plotted in Fig.~\ref{Fig15} and it appears that for $St>0. 45$ strong one-dimensional clustering has vanished.

\begin{figure}[!ht]
\includegraphics [width=1.\columnwidth,height=0.75\columnwidth]{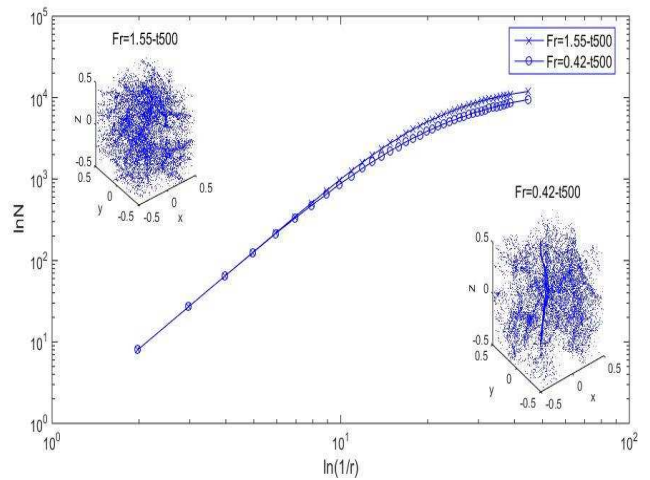}
\caption{\label{Fig16}Box counting slope for very similar cases for different values of $Fr$ at $St=0.207$}
\end{figure}

A major problem with the box counting method is to discriminate between the dimension $D$ of very similar patterns as shown in Fig.~\ref{Fig16}.Therefore, we cannot be sure that the box-counting result is meaningful in a case with no clear structures (i.e. without an integer dimension. As mentioned earlier, in order to be accurate, the BCM must be applied to the Lagrangian attractor. If the particle  cloud has not settled on the attractor as at time $t=10$ or $t=600$ in Fig.~\ref{Fig2}, the BCM will not educe the 1D patterns. So, in practice, it means running the cases for long times until the particles have all settled on the Lagrangian attractor which can be prohibitive.

\subsection{Nearest-neighbour distance analysis}
The advantage of using this approach is that it is not necessary to reach the final cluster at $t \to \infty$, a snapshot at earlier times gives us a clear idea of
the kind of Lagrangian attractor to expect. For example in Fig.~\ref{Fig2} the kind of 1D-V Lagrangian attractor is clear at $t=10$ there is no need to wait
for the final asymptotic shape at $t=1200$. Though other methods may struggle to pick up the structure at $t=10$ and will only give the correct diagnosis when
all the points have settle on the attractor, that is for $t=1200$, the nearest-neighbour distance analysis will work for intermediary times.
The average-distance-to-the-nearest-neighbour $\Delta$ \cite{Park-Lee-2014}
is introduced to systematically quantify the clustering patterns.
At a given time for each particle $\mathbf{X}_m$ its nearest neighbour $\mathbf{X}_n$ is such that
\begin{equation}
\Delta^2_{mi}=(x_{m}-x_{i})^{2}+(y_{m}-y_{i})^{2}+( z_{m}-z_{i})^{2}
\label{E2-31c}
\end{equation}
is a minimum for $i=n$. Then we define the average-distance-to-the-nearest-neighbour as
\begin{equation}
\Delta=\frac{1}{N_p}\sqrt{\sum_{m=1}^{N_p}\Delta^2_{mn}}
\label{E2-31d}
\end{equation}
Where $\mathbf{X}_n=(x_{n},y_{n},z_{n})$ is the nearest particle's neighbour.

We get three obvious benchmark values for this method.
\begin{itemize}
\item[i)] If the particles are homogeneously distributed as at time $t=0$, (Fig.~\ref{fig-01-partpos}a) then $\Delta \simeq L_x/N = 1/25 = 0.04$.
\item[ii)] If the particles are distributed on a surface-like attractor 2D-L as in Fig.~\ref{Fig11}) then $\Delta \simeq L_x/N^{3/2} = 1/25^{3/2} = 0.008$.
\item[iii)] If the particles are distributed on a line-like attractor as in Fig.~\ref{Fig2}) then $\Delta \simeq L_x/N^3 = 1/25^3 = 6.4\, 10^{-5}$.
\end{itemize}

\begin{figure}[htp!]
\includegraphics [width=1.\columnwidth,height=0.75\columnwidth]{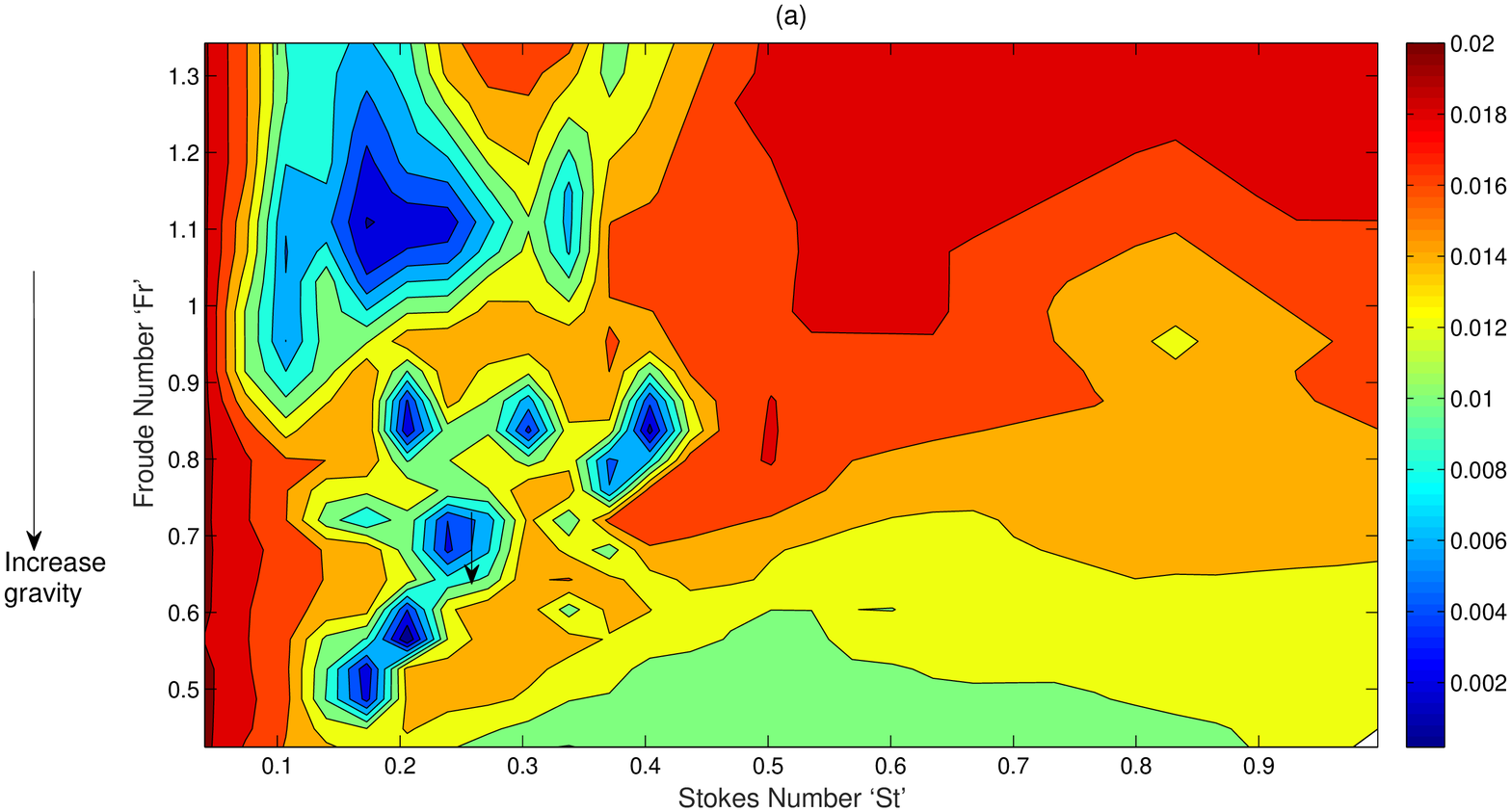}
\includegraphics [width=1.\columnwidth,height=0.75\columnwidth]{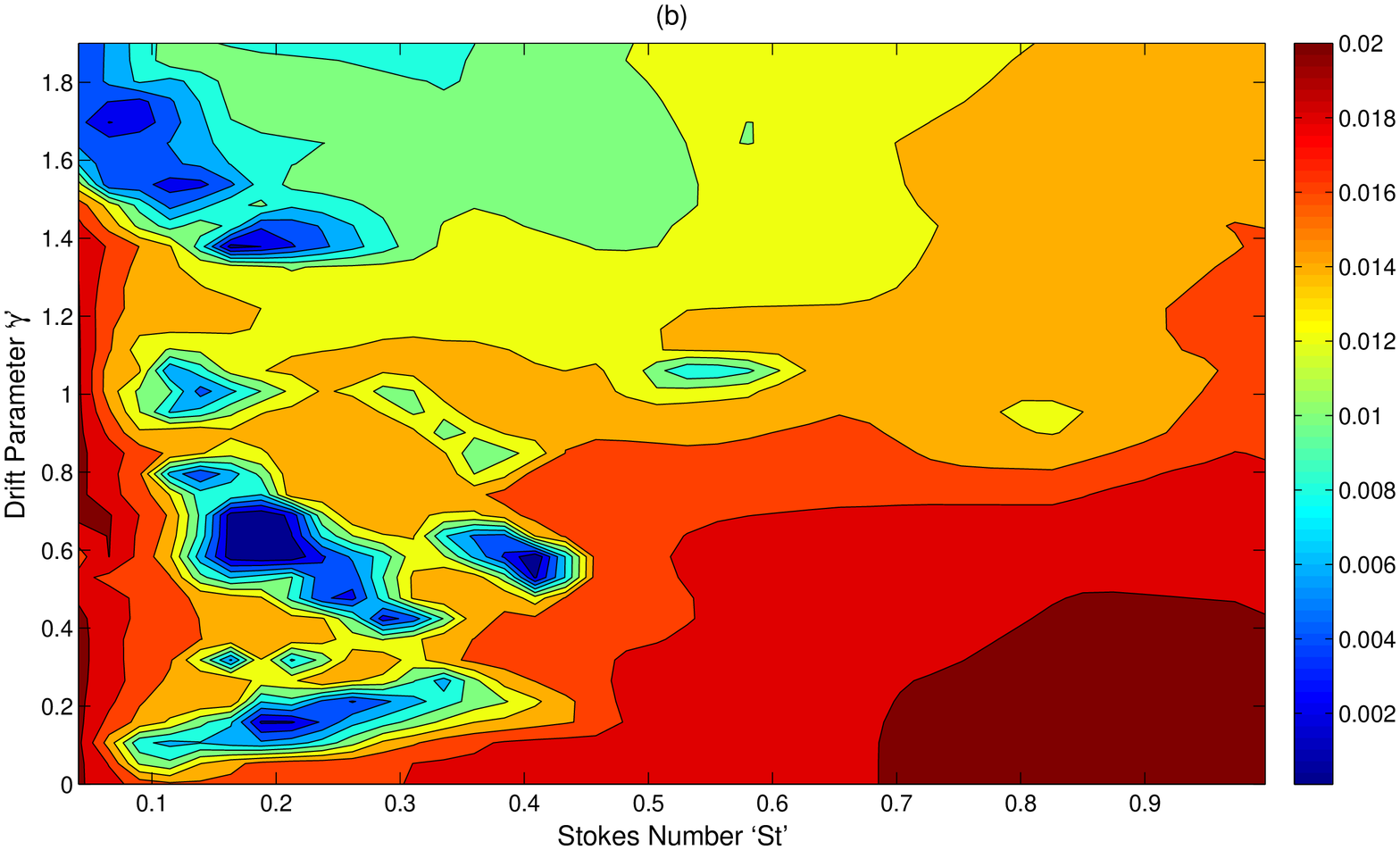}
\caption{\label{Fig19} (Color
  online) Iso-contours of: a) $\Delta(St,Fr)$ and b) $\Delta(St,\gamma)$ showing different types of clusterings}
\end{figure}

In practice, the method will detect a one-dimensional structure (1D-H or 1-DV) for $\Delta \le 0.008$ while 2D layered structures are observed for $0.01 \le \Delta \le 0.014$.
\\[2ex]
We applied the average-distance-to-nearest-neighbour method to all run cases to see the variations in the attractor patterns
for the same time $t=300$. Colour-wise blue (darkest spots for $St \le 0.4$) corresponds to the 1D Lagrangian attractor, yellow-green (light grey) to  the 2D-L and dark red (darkest spots for $St \ge 0.4$) to 3D structures. Iso-contours of $\Delta$ are plotted as functions of ($St$, $Fr$) in Fig.~\ref{Fig19}a and ($St$, $\gamma$) in Fig.~\ref{Fig19}b to see the effect of varying gravity and inertia on the Lagrangian attractors.
Fig.~\ref{Fig19} confirms that the clustering of inertial particles is not a monotonic function of either $St$ or $Fr$ number. However, it is possible to identify regions in the plane $(St, Fr)$:
\begin{itemize}
\item[i)]
In agreement with Fig.~\ref{Fig15}, there is no one-dimensional structures for $St \ge 0.5$.
\item[ii)]
We can refine the analysis in term of $\gamma$:
In Fig.~\ref{Fig19}b, for large values of the Stokes number ($St > 0.3$) and gravity effects ($\gamma > 0.8$) the 2D-L structures are predominant. This is also in agreement with \cite{Gustavsson-al-2014} whose calculations show that for large values of $St$
particles may cluster strongly.
\end{itemize}
Another advantage of the average-distance-to-nearest-neighbour method is that variations in horizontal $\Delta_{H}$ and vertical $\Delta_{V}$ directions can be identified separately which help to monitor anisotropic patterns.
In practice, $\Delta_{H}$ and $\Delta_{V}$ are defined as follows:
\begin{equation}
\Delta_{H}=\frac{1}{N_p}\sqrt{\sum_{m=1}^{N_p} ( x_{m}-x_{n})^{2}+( y_{m}-y_{n})^{2}}
        \label{E2-31e}
\end{equation}
\begin{equation}
\Delta_{V}=\frac{1}{N_p}\sqrt{\sum_{m=1}^{N_p}( z_{m}-z_{n})^{2}}
        \label{E2-31f}
\end{equation}
\begin{figure*}[!ht]
\includegraphics [width=2.\columnwidth]{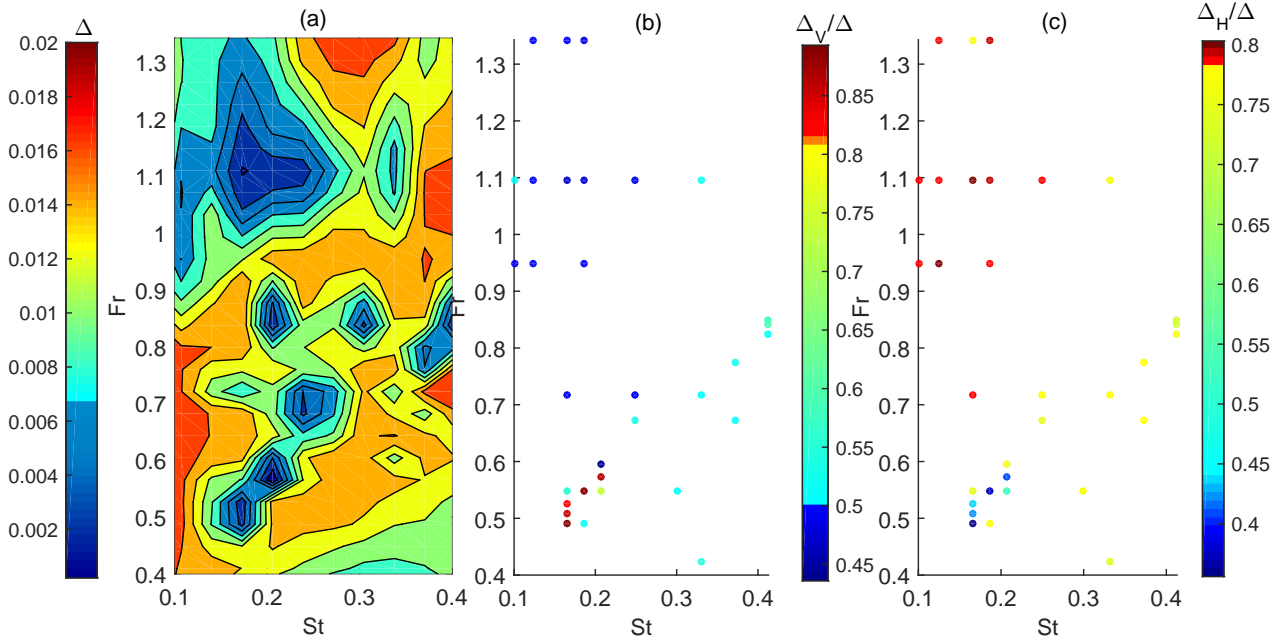}
\caption{\label{Fig20} (Color
  online) a) Iso-contours of $\Delta$ as a function of ($St$, $Fr$),
b) $\Delta_V/\Delta$ when $\Delta \le 0.006$, c) $\Delta_H/\Delta$ when $\Delta \le 0.006$.
}
\end{figure*}
Fig.~\ref{Fig20}a shows the iso-contours of $\Delta$ as a function of ($St$, $Fr$) in the region where one-dimensional Lagrangian attractors are observed, i.e. $0.05 \le St \le 0.4$.
The two different types of one-dimensional attractor either horizontal (1D-H) or vertical (1D-V) can be further analysed in Figs~\ref{Fig20}b and c.
Fig.~\ref{Fig20}c shows the ratio $\Delta_H/\Delta$ where one-dimensional attractors exist, that is when $\Delta \le 0.008$. $\Delta_H/\Delta \le 0.5$, that is blueish points (light grey area), indicates 1D-V structures; whereas $\Delta_H/\Delta \ge 0.75$, that is redish points (dark spots for $Fr \ge 0.7$), indicates 1D-H structures.
Fig.~\ref{Fig20}b describes a similar relationship based on $\Delta_V$. So it is clear from the points' colours distribution that horizontal attractors are predominant for large $Fr$ while vertical attractors are prevalent as $Fr$ decreased.
%

\section{Conclusion}
\label{seconcl}
We used Kinematic Simulation (KS) to study the clustering pattern of particles with inertia
subjected to gravity effects. For some combined inertia and gravity effects ($St$, $Fr$), the particles cluster on a fixed space subset. That subset can be one-dimensional or two-dimensional. In most of the cases, the particles did not cluster and disperse occupying the most of the periodic box.

Using KS, it became possible to investigate many combinations of ($St$, $Fr$) and educe and classify those one-dimensional or two-dimensional subsets. Though KS retains only part of the turbulence physics, it helps to understand the clustering patterns and the effect of gravity on these patterns.


\begin{figure*}[!ht]
\includegraphics [width=2.\columnwidth]{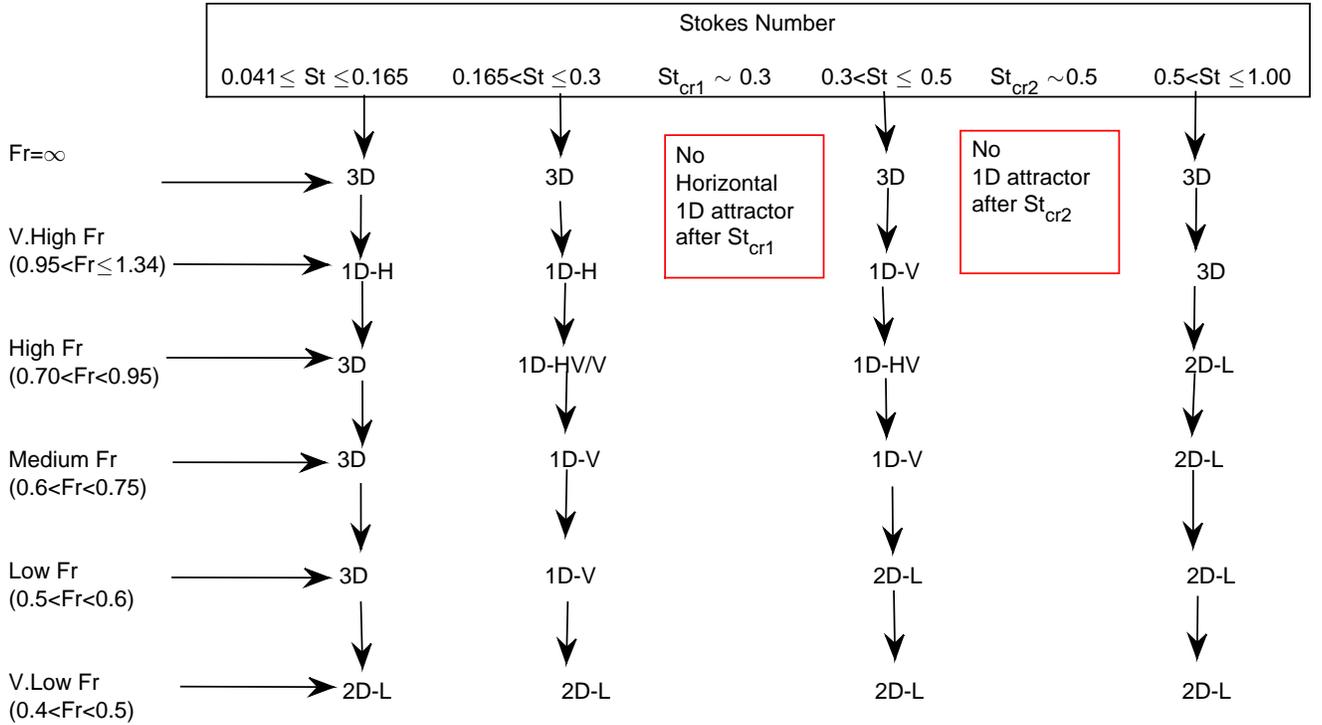}
 \caption{\label{Fig11b} Flow chart describing the different attractors in relation to the two critical values of $St$}
\end{figure*}
The main results can be summarised as follows (and in a more synthetic presentation in Figure~\ref{Fig11b}):
\begin{itemize}
\item
The effect of gravity may reduce or enhance inertial particles clustering (as noticed in \cite{Gustavsson-al-2014,Bec-et-al-2014}) depending on the Stokes number. This effect can lead to strongly anisotropic clusterings (1D or 2D-L) very clearly evidenced by the KS model.
\item
The 1D structure is better observed with the synthetic flow, as in real flows unsteadiness may prevent the particles from reaching that asymptotic state.
These 1D attractors move from the horizontal to the vertical direction as the $Fr$ number decreases.
\item
For our range of Froude numbers, we found two critical Stokes numbers: for $St > St_{cr1}=0.3$
there is never occurrence of a horizontal (1D-H) type attractor and no 1D-type attractor is found for $St > St_{cr2}=0.5$.
\item
For low values of $Fr$, curtain-like two-dimensional layered structures
similar to the `curtain-like manifolds'. already observed in \cite{Woittiez-et-al-2008}
are recovered as the high gravity prevents the inertial particles from settling uniformly in the turbulent flow.
\end{itemize}

\section{Acknowledgement}

M. Farhan gratefully acknowledges support funding from the Department of Mechanical Engineering of the University of Engineering and Technology Lahore, Pakistan.
\\
This work was supported by EPSRC grant EP/L000261/1



\begin{thebibliography}{21}
\expandafter\ifx\csname natexlab\endcsname\relax\def\natexlab#1{#1}\fi
\expandafter\ifx\csname bibnamefont\endcsname\relax
  \def\bibnamefont#1{#1}\fi
\expandafter\ifx\csname bibfnamefont\endcsname\relax
  \def\bibfnamefont#1{#1}\fi
\expandafter\ifx\csname citenamefont\endcsname\relax
  \def\citenamefont#1{#1}\fi
\expandafter\ifx\csname url\endcsname\relax
  \def\url#1{\texttt{#1}}\fi
\expandafter\ifx\csname urlprefix\endcsname\relax\def\urlprefix{URL }\fi
\providecommand{\bibinfo}[2]{#2}
\providecommand{\eprint}[2][]{\url{#2}}
\bibitem[{\citenamefont{Falkovich et~al.}(2002)\citenamefont{Falkovich, Fouxon,
  and Stepanov}}]{Falkovich-al-2002}
\bibinfo{author}{\bibfnamefont{G.}~\bibnamefont{Falkovich}},
  \bibinfo{author}{\bibfnamefont{A.}~\bibnamefont{Fouxon}}, \bibnamefont{and}
  \bibinfo{author}{\bibfnamefont{M.~G.}~\bibnamefont{Stepanov}},
  \bibinfo{journal}{Nature} \textbf{\bibinfo{volume}{419}},
  \bibinfo{pages}{151} (\bibinfo{year}{2002}).
\bibitem[{\citenamefont{Pan et~al.}(2011)\citenamefont{Pan,Padoan, Scalo, Kritsuk and
  Norman}}]{Pan-al-2011}
\bibinfo{author}{\bibfnamefont{L.}~\bibnamefont{Pan}},
  \bibinfo{author}{\bibfnamefont{P.}~\bibnamefont{Padoan}},
  \bibinfo{author}{\bibfnamefont{J.}~\bibnamefont{Scalo}},
  \bibinfo{author}{\bibfnamefont{A.G.}~\bibnamefont{Kritsuk}}, \bibnamefont{and}
  \bibinfo{author}{\bibfnamefont{M.L.}~\bibnamefont{Norman}},
  \bibinfo{journal}{The Astrophysical Journal} \textbf{\bibinfo{volume}{740}},
  \bibinfo{pages}{21} (\bibinfo{year}{2011}).
\bibitem[{\citenamefont{Cencini et~al.}(2006)\citenamefont{Cencini,Bec,Biferale,Boffetta,Celani,Lanotte,Musacchio and Toschi}}]{Cencini-al-2006}
\bibinfo{author}{\bibfnamefont{M.}~\bibnamefont{Cencini}},
  \bibinfo{author}{\bibfnamefont{J.}~\bibnamefont{Bec}},
  \bibinfo{author}{\bibfnamefont{L.}~\bibnamefont{Biferale}},
  \bibinfo{author}{\bibfnamefont{G.}~\bibnamefont{Boffetta}},
  \bibinfo{author}{\bibfnamefont{A.}~\bibnamefont{Celani}},
  \bibinfo{author}{\bibfnamefont{A.}~\bibnamefont{Lanotte}},
  \bibinfo{author}{\bibfnamefont{S.}~\bibnamefont{Musacchio}} \bibnamefont{and}
  \bibinfo{author}{\bibfnamefont{F.} \bibnamefont{Toschi}},
  \bibinfo{journal}{Journal of Turbulence} \textbf{\bibinfo{volume}{7}},
  \bibinfo{pages}{36} (\bibinfo{year}{2006}).
\bibitem[{\citenamefont{Saw et~al.}(2006)\citenamefont{Saw, Salazar, Collins, and
  Shaw}}]{Saw-al-2012}
\bibinfo{author}{\bibfnamefont{E.W.}~\bibnamefont{Saw}},
  \bibinfo{author}{\bibfnamefont{J.P.L.C.}~\bibnamefont{Salazar}},
  \bibinfo{author}{\bibfnamefont{L.R.}~\bibnamefont{Collins}}, \bibnamefont{and}
  \bibinfo{author}{\bibfnamefont{R.A.}~\bibnamefont{Shaw}},
  \bibinfo{journal}{New Journal of Physics} \textbf{\bibinfo{volume}{14}},
  \bibinfo{pages}{105030} (\bibinfo{year}{2012}).

\bibitem[{\citenamefont{Falkovich-Pumir}(2004)\citenamefont{Falkovich and Pumir}}]{Falkovich-Pumir-2004}
  \bibinfo{author}{\bibfnamefont{G.}~\bibnamefont{Falkovich}}, \bibnamefont{and}
  \bibinfo{author}{\bibfnamefont{A.}~\bibnamefont{Pumir}},
  \bibinfo{journal}{Phys. Fluids} \textbf{\bibinfo{volume}{16}},
  \bibinfo{pages}{L47} (\bibinfo{year}{2004}).

\bibitem[{\citenamefont{Bec et~al.}(2007)\citenamefont{Bec,Biferale,Cencini,Lanotte,Musacchio and Toschi}}]{Bec-al-2007}
  \bibinfo{author}{\bibfnamefont{J.}~\bibnamefont{Bec}},
  \bibinfo{author}{\bibfnamefont{L.}~\bibnamefont{Biferale}},
  \bibinfo{author}{\bibfnamefont{M.}~\bibnamefont{Cencini}},
  \bibinfo{author}{\bibfnamefont{A.}~\bibnamefont{Lanotte}},
  \bibinfo{author}{\bibfnamefont{S.}~\bibnamefont{Musacchio}}~\bibnamefont{and}
  \bibinfo{author}{\bibfnamefont{F.}~\bibnamefont{Toschi}},
  \bibinfo{journal}{Phys. Rev. Lett.} \textbf{\bibinfo{volume}{98}},
  \bibinfo{pages}{084502} (\bibinfo{year}{2007}).

\bibitem[{\citenamefont{Fung et~al.}(1992)\citenamefont{Fung, Hunt, Malik, and
  Perkins}}]{Fung-al-1992}
\bibinfo{author}{\bibfnamefont{J.}~\bibnamefont{Fung}},
  \bibinfo{author}{\bibfnamefont{J.}~\bibnamefont{Hunt}},
  \bibinfo{author}{\bibfnamefont{N.}~\bibnamefont{Malik}}, \bibnamefont{and}
  \bibinfo{author}{\bibfnamefont{R.}~\bibnamefont{Perkins}},
  \bibinfo{journal}{J. Fluid Mech.} \textbf{\bibinfo{volume}{236}},
  \bibinfo{pages}{281} (\bibinfo{year}{1992}).

\bibitem[{\citenamefont{Elliott and Majda}(1996)}]{Elliott-Majda1996}
\bibinfo{author}{\bibfnamefont{F.~W.} \bibnamefont{Elliott}} \bibnamefont{and}
  \bibinfo{author}{\bibfnamefont{A.~J.} \bibnamefont{Majda}},
  \bibinfo{journal}{Phys. Fluids} \textbf{\bibinfo{volume}{8}},
  \bibinfo{pages}{1052} (\bibinfo{year}{1996}).

\bibitem[{\citenamefont{Malik and Vassilicos}(1999)}]{Malik-Vassilicos1999}
\bibinfo{author}{\bibfnamefont{N.~A.} \bibnamefont{Malik}} \bibnamefont{and}
  \bibinfo{author}{\bibfnamefont{J.~C.} \bibnamefont{Vassilicos}},
  \bibinfo{journal}{Phys. Fluids} \textbf{\bibinfo{volume}{11}},
  \bibinfo{pages}{1572} (\bibinfo{year}{1999}).

\bibitem[{\citenamefont{Nicolleau and Nowakowski}(2011)}]{Nicolleau-Nowakowski-2011}
\bibinfo{author}{\bibfnamefont{F.C.G.A}~\bibnamefont{Nicolleau}} \bibnamefont{and}
\bibinfo{author}{\bibfnamefont{A.F.}~\bibnamefont{Nowakowski}},
\bibinfo{journal}{Phys. Rev. E} \textbf{\bibinfo{volume}{83}},
\bibinfo{pages}{056317} (\bibinfo{year}{2011}).

\bibitem[{\citenamefont{Malik}(2014{\natexlab{a}})}]{Malik-2014a}
\bibinfo{author}{\bibfnamefont{N.~A.} \bibnamefont{Malik}},
  \bibinfo{journal}{http://arxiv.org/abs/1405.3625.}
  (\bibinfo{year}{2014}{\natexlab{a}}).

\bibitem[{\citenamefont{Malik}(2014{\natexlab{b}})}]{Malik-2014b}
\bibinfo{author}{\bibfnamefont{N.~A.} \bibnamefont{Malik}},
  \bibinfo{journal}{http://arxiv.org/abs/1405.3638.}
  (\bibinfo{year}{2014}{\natexlab{b}}).

\bibitem[{\citenamefont{Nicolleau and Yu}(2007)}]{Nicolleau-Yu-2007}
\bibinfo{author}{\bibfnamefont{F.}~\bibnamefont{Nicolleau}} \bibnamefont{and}
  \bibinfo{author}{\bibfnamefont{G.}~\bibnamefont{Yu}}, \bibinfo{journal}{Phys.
  Rev. E} \textbf{\bibinfo{volume}{76}}, \bibinfo{pages}{066302}
  (\bibinfo{year}{2007}).

\bibitem[{\citenamefont{Nicolleau et~al.}(2013)\citenamefont{Nicolleau, Sung,
  and Vassilicos}}]{Nicolleau-al-2012-ftac}
\bibinfo{author}{\bibfnamefont{F.}~\bibnamefont{Nicolleau}},
  \bibinfo{author}{\bibfnamefont{K.-S.} \bibnamefont{Sung}}, \bibnamefont{and}
  \bibinfo{author}{\bibfnamefont{J.}~\bibnamefont{Vassilicos}},
  \bibinfo{journal}{Flow, Turbulence and Comb.} \textbf{\bibinfo{volume}{91}},
  \bibinfo{pages}{79} (\bibinfo{year}{2013}).

\bibitem[{\citenamefont{Ijzermans et~al.}(2010)\citenamefont{Ijzermans,
  Meneguz, and Reek}}]{Izermans-et-al-2010}
\bibinfo{author}{\bibfnamefont{R.~H.~A.} \bibnamefont{Ijzermans}},
  \bibinfo{author}{\bibfnamefont{E.}~\bibnamefont{Meneguz}}, \bibnamefont{and}
  \bibinfo{author}{\bibfnamefont{M.~W.} \bibnamefont{Reek}},
  \bibinfo{journal}{J. Fluid Mech.} \textbf{\bibinfo{volume}{653}},
  \bibinfo{pages}{99–136} (\bibinfo{year}{2010}).

\bibitem[{\citenamefont{Meneguz and Reeks}(2011)}]{Meneguz-Reeks-2011}
\bibinfo{author}{\bibfnamefont{E.}~\bibnamefont{Meneguz}} \bibnamefont{and}
  \bibinfo{author}{\bibfnamefont{M.~W.} \bibnamefont{Reeks}},
  \bibinfo{journal}{J. Fluid Mech.} \textbf{\bibinfo{volume}{686}},
  \bibinfo{pages}{338} (\bibinfo{year}{2011}).

\bibitem[{\citenamefont{Fung and Vassilicos}(1998)}]{Fung-Vassilicos-1998}
\bibinfo{author}{\bibfnamefont{J.C.H.}~\bibnamefont{Fung}} \bibnamefont{and}
  \bibinfo{author}{\bibfnamefont{J.C.}~\bibnamefont{Vassilicos}},
  \bibinfo{journal}{Phys. Rev. E} \textbf{\bibinfo{volume}{57}},
  \bibinfo{pages}{1677} (\bibinfo{year}{1998}).

\bibitem[{\citenamefont{Nicolleau and ElMaihy}(2006)}]{Nicolleau-ElMaihy-2006}
\bibinfo{author}{\bibfnamefont{F.}~\bibnamefont{Nicolleau}} \bibnamefont{and}
  \bibinfo{author}{\bibfnamefont{A.}~\bibnamefont{ElMaihy}},
  \bibinfo{journal}{Phys Rev. E} \textbf{\bibinfo{volume}{74}},
  \bibinfo{pages}{046302} (\bibinfo{year}{2006}).

\bibitem[{\citenamefont{Abou-El-Azm and
  Nicolleau}(2008)}]{El-Azm-Nicolleau-2008}
\bibinfo{author}{\bibfnamefont{A.}~\bibnamefont{Abou-El-Azm}} \bibnamefont{and}
  \bibinfo{author}{\bibfnamefont{F.}~\bibnamefont{Nicolleau}},
  \bibinfo{journal}{Phys. Rev. E} \textbf{\bibinfo{volume}{78}},
  \bibinfo{pages}{0616310} (\bibinfo{year}{2008}).

\bibitem[{\citenamefont{Gatignol}(1983)}]{Gatignol-1983}
\bibinfo{author}{\bibfnamefont{R.}~\bibnamefont{Gatignol}},
  \bibinfo{journal}{J. Mech. Theor. Appl.} \textbf{\bibinfo{volume}{1}},
  \bibinfo{pages}{143} (\bibinfo{year}{1983}).

\bibitem[{\citenamefont{Maxey and Riley}(1983)}]{Maxey-Riley-1983}
\bibinfo{author}{\bibfnamefont{M.~R.} \bibnamefont{Maxey}} \bibnamefont{and}
  \bibinfo{author}{\bibfnamefont{J.~J.} \bibnamefont{Riley}},
  \bibinfo{journal}{Phys. Fluids} \textbf{\bibinfo{volume}{26}},
  \bibinfo{pages}{883} (\bibinfo{year}{1983}).

\bibitem[{\citenamefont{Woittiez et~al.}(2008)\citenamefont{Woittiez, Jonker,
  and nd~Lu\'{\i}s M.~Portela}}]{Woittiez-et-al-2008}
\bibinfo{author}{\bibfnamefont{E.~J.~P.} \bibnamefont{Woittiez}},
  \bibinfo{author}{\bibfnamefont{H.~J.~J.} \bibnamefont{Jonker}},
  \bibnamefont{and} \bibinfo{author}{\bibnamefont{L. M.~Portela}},
  \bibinfo{journal}{Journal of the Atmospheric Sciences}
  \textbf{\bibinfo{volume}{66}}, \bibinfo{pages}{1926} (\bibinfo{year}{2008}).

\bibitem[{\citenamefont{Yongnam~Park}(2014)}]{Park-Lee-2014}
\bibinfo{author}{\bibfnamefont{Y.} \bibnamefont{Park}},
\bibinfo{author}{\bibfnamefont{C.} \bibnamefont{Lee}},
  \bibinfo{journal}{Phys. Rev. E} \textbf{\bibinfo{volume}{89}},
  \bibinfo{pages}{061004(R)} (\bibinfo{year}{2014}).

\bibitem[{\citenamefont{Nicolleau and ElMaihy}(2004)}]{Nicolleau-ElMaihy-2004}
\bibinfo{author}{\bibfnamefont{F.}~\bibnamefont{Nicolleau}} \bibnamefont{and}
  \bibinfo{author}{\bibfnamefont{A.}~\bibnamefont{ElMaihy}},
  \bibinfo{journal}{J. Fluid Mech.} \textbf{\bibinfo{volume}{517}},
  \bibinfo{pages}{229} (\bibinfo{year}{2004}).

\bibitem[{\citenamefont{Gustavsson et~al.}(2014)\citenamefont{Gustavsson,
  Vajedi, and Mehlig}}]{Gustavsson-al-2014}
\bibinfo{author}{\bibfnamefont{K.}~\bibnamefont{Gustavsson}},
  \bibinfo{author}{\bibfnamefont{S.}~\bibnamefont{Vajedi}}, \bibnamefont{and}
  \bibinfo{author}{\bibfnamefont{B.}~\bibnamefont{Mehlig}},
  \bibinfo{journal}{Phys. Rev. Lett.} \textbf{\bibinfo{volume}{112}},
  \bibinfo{pages}{214501} (\bibinfo{year}{2014}).

\bibitem[{\citenamefont{Bec et~al.}(2014)\citenamefont{Bec, Homann, and
  Ray}}]{Bec-et-al-2014}
\bibinfo{author}{\bibfnamefont{J.}~\bibnamefont{Bec}},
  \bibinfo{author}{\bibfnamefont{H.}~\bibnamefont{Homann}}, \bibnamefont{and}
  \bibinfo{author}{\bibfnamefont{S.~S.} \bibnamefont{Ray}},
  \bibinfo{journal}{Phys. Rev. Lett.} \textbf{\bibinfo{volume}{112}},
  \bibinfo{pages}{184501} (\bibinfo{year}{2014}).

\end{thebibliography}
\bibliographystyle{apsrev}

\end{document}